\documentclass[useAMS,usenatbib]{mn2e} 
\usepackage{amsmath}
\usepackage{color}
\usepackage{epsf}
\usepackage{epsfig}
\usepackage{ulem}
\usepackage{graphicx}
\usepackage{natbib}
\usepackage{textcomp}
\newif\ifAMStwofonts
\bibpunct{(}{)}{;}{a}{}{,}
\usepackage{graphicx,amsmath,amssymb,epsfig}
\usepackage{hyperref}
\RequirePackage{lineno}
\usepackage{astrobib_mnras2e} 
\newcommand{\zdrag}{z_{\rm drag}}
\newcommand{\Om}{\Omega_m}
\newcommand{\Ok}{\Omega_K}
\newcommand{\Ob}{\Omega_b}
\newcommand{\LCDM}{\rm {\Lambda CDM}}

\newcommand{\Nb}{{\it N}-body}
\newcommand{\Nrel}{N_{\rm REL}}

\title[A 2\% Distance to z=0.35: Cosmological Measurements]
{A 2\% Distance to z = 0.35 by Reconstructing Baryon Acoustic Oscillations - III : Cosmological Measurements and Interpretation}
\author[K.~Mehta et al.]
{Kushal T. Mehta$^1$, Antonio J. Cuesta$^2$, Xiaoying Xu$^1$, Daniel J. Eisenstein$^3$,
\newauthor 
Nikhil Padmanabhan$^2$ \\
$^1$Steward Observatory, University of Arizona, 
933 N. Cherry Ave., Tucson, AZ 85121\\
$^2$Dept. of Physics, Yale University,
260 Whitney Avenue, New Haven, CT 06511\\
$^3$Harvard-Smithsonian Center for Astrophysics, Harvard University,
60 Garden St., Cambridge, MA 02138}

\begin{document}

\maketitle

\begin{abstract}
We use the $2\%$ distance measurement from our reconstructed baryon acoustic oscillations (BAOs) signature using the Sloan Digital Sky Survey (SDSS) Data Release 7 (DR7) Luminous Red Galaxies (LRGs) from \citet{paperI} and \citet{paperII} combined with cosmic microwave background (CMB) data from Wilkinson Microwave Anisotropy Probe (WMAP7) to measure parameters for various cosmological models. We find a $1.7\%$ measurement of $H_0 = 69.8 \pm 1.2$ km s$^{-1}$ Mpc$^{-1}$ and a $5.0\%$ measurement of $\Om = 0.280 \pm 0.014$ for a flat Universe with a cosmological constant. These measurements of $H_0$ and $\Om$ are robust against a range of underlying models for the expansion history. We measure the dark energy equation of state parameter $w = -0.97 \pm 0.17$, which is consistent with a cosmological constant. If curvature is allowed to vary, we find that the Universe is consistent with a flat geometry ($\Omega_K = -0.004 \pm 0.005$). We also use a combination of the 6 Degree Field Galaxy Survey BAO data, WiggleZ Dark Energy Survey data, Type Ia supernovae (SN) data, and a local measurement of the Hubble constant to explore cosmological models with more parameters. Finally, we explore the effect of varying the energy density of relativistic particles on the measurement of $H_0$.
\end{abstract}

\begin{keywords}
distance scale 
--- cosmological parameters
--- large-scale structure of Universe
--- cosmology: theory, observations
\end{keywords}

\section{Introduction}\label{sec:intro}
Since the discovery of the accelerated expansion of the Universe \citep{riess98, perlmutter99}, there has been a growing interest to understand the nature of dark energy and measure various cosmological parameters. This understanding requires improved measurements of the expansion history of the Universe via the distance-redshift relation. In particular, baryon acoustic oscillations (BAO) have been widely used to study this relation by measuring cosmic distances. The physics behind these oscillations is well understood \citep{sakharov67, peebles70, sunyaev70, bond84, bond87, hu96a, hu97, EH98, hu02}. In the pre-recombination era of the Universe, the baryons were coupled to the photons in a hot plasma. Small overdensities in the underlying dark matter distribution caused the baryons and photons to fall into the overdensities due to gravity. As the plasma density grows, the radiation pressure from the photons drive an acoustic wave of baryons and photons around the original dark matter overdensity. As the Universe cools, the electrons and protons combine to form atoms, and the photons decouple from the baryons causing the sound speed in the plasma to drop dramatically. This leaves the baryons in a spherical shell around the initial overdensity. This shell has a characteristic scale of about 150 Mpc, defined by the distance traveled by the acoustic wave in the pre-recombination era, and its angular scale has been measured in the cosmic microwave background (CMB) to be about $1^\circ$ \citep{bennett03, jarosik11}. Under the influence of gravity, these overdensities grow and form galaxies imprinting the characteristic acoustic scale into the distribution of galaxies \citep{hu96a, EH98, meiksin99}. Thus, the BAO scale can be used as a robust standard ruler in large galaxy surveys \citep{tegmark97, goldberg98, eisenstein98a, efstathiou99} with an important application to the study of dark energy \citep{eisenstein02, blake03, hu03, linder03, seo03}. The large physical size of this acoustic scale causes the standard ruler to be highly accurate \citep{eisenstein07, seo08, padmanabhan09a, seo10, mehta11}.

The BAO signal was first measured in the SDSS LRG survey and the 2dF Galaxy survey \citep{eisenstein05, cole05} and has since been observed in multiple surveys SDSS \citep{tegmark06, percival07, percival10, kazin10, chuang10}, 6dF Galaxy Survey \citep{beutler11}, and the WiggleZ Dark Energy Survey \citep{blake10, blake11a, blake11b}. \citet{weinberg12} provides an overall review of observational cosmology and discusses the current state of the field in depth. 

The acoustic scale hence gives us a measurement of the distance to a given redshift. \cite{paperI} (hereafter Paper I) presents the BAO measurements via the correlation function in the SDSS Luminous Red Galaxies (LRG) Data Release 7 (DR7) dataset using the reconstruction technique first introduced by \cite{eisenstein07a}. \cite{paperII} (hereafter PaperII) describes a robust methodology to measure the acoustic scale, which is heavily tested against the LasDamas mock catalogs. Also shown in PaperII are the results of the testing performance of the reconstruction technique to improve the BAO measurement. The reconstruction technique improves the distance measurement to $z = 0.35$ to a $1.9\%$ measurement compared to a $3.5\%$ measurement before reconstruction. We show in this paper how this new measurement of the acoustic scale in SDSS helps improve our measurements of the cosmological parameters over a wide range of cosmological models. 

The CMB angular acoustic scale gives us a distance measurement to the redshift at recombination that helps us break the degeneracy between $\Om$ and $H_0$, therefore precisely measuring the parameters in the flat $\LCDM$ or ``vanilla'' cosmological model. However, with higher dimensional models, we need to have more distance measurements to break degeneracies between various cosmological parameters. We show how BAO data helps break these degeneracies by providing a second distance measurement at low redshift. We extend our redshift range to lower redshifts by adding SN data.   

The combination of degree-scale CMB anisotropy, large-scale structure, and SN Ia data offers powerful constraints on cosmology and dark energy. Notable early papers include \citet{efstathiou02}, \citet{percival02}, \citet{spergel03}, and \citet{tegmark04}. With the discovery of the acoustic peak in the large-scale clustering of galaxies, the results from the WMAP satellite, and the construction of yet-larger supernova samples, these constraints have gotten increasingly precise. Many papers have combined these data sets; some recent examples include \citet{komatsu09}, \citet{hicken09}, \citet{kazin10}, \citet{percival10}, \citet{reid10a}, \citet{blake10}, \citet{komatsu11}, \citet{conley11}, \citet{blake11a}, \citet{wang11}, \citet{beutler11}, \citet{seo12}, and \citet{ho12} (see \citet{weinberg12} for a longer discussion).

In this paper, we use the results of this reconstructed SDSS data in conjunction with CMB measurements from the Seven-year Wilkinson Microwave Anisotropy Probe (WMAP7)~\citep{jarosik11, komatsu11}, Type Ia supernovae (SN) measurements from the 3 year Supernovae Legacy Survey (SNLS) \citep{conley11}, and direct measurement of the Hubble constant from the SH0ES project  \citep{riess11}. To break degeneracies between different parameters, we use additional BAO data from the 6dF Galaxy Survey \citep{beutler11}, the WiggleZ Dark Energy Survey \citep{blake11}, and SN data from the 3 year Supernova Legacy Survey (SNLS3) by \cite{conley11}. 

We start with the concordance cosmological model, $\LCDM$ (which here denotes a flat Universe), and add other cosmological parameters to explore higher dimensional models. We vary the curvature of the Universe, $\Ok$, and the constant dark energy equation of state, $w_0$ independently for the oCDM and $w$CDM models respectively. For higher dimensionality, we vary both $\Ok$ and $w_0$ simultaneously in the o$w$CDM model and in the $w_0w_a$CDM model we assume a flat Universe but allow the dark energy equation of state parameter to vary in time. We allow all three parameters: $\Ok, w_0, w_a$ to vary in our most general model, o$w_0w_a$CDM. 

In Section~\ref{sec:methods}, we describe the Markov Chain Monte Carlo fitting and the various datasets we used in this study. We introduce the results of the BAO data and describe the cosmological implications in Section~\ref{sec:ladder}. In Sections~\ref{sec:lcdm} to \ref{sec:ow0wacdm}, we show our results for various cosmological models. In Section~\ref{sec:H0Omegam} we show the robustness of our measurements of Hubble constant and the matter density over different models for the expansion history. Section~\ref{sec:nrel} explores a possibility to solve an apparent tension in the $H_0$ measurement by varying the energy density of relativistic species. We conclude with a summary of our results in Section~\ref{sec:conclusions}.

\section{Methodology}\label{sec:methods}
Here, in the third paper of this series, we use the reconstructed SDSS DR7 LRG results presented in PaperI and PaperII to measure cosmological parameters. We use a reconstruction technique first introduced in \cite{eisenstein07a}, to model and remove effects of non-linear evolution of large scale structure and large scale velocity flows (redshift space distortions). As shown in \cite{mehta11}, this method can also be applied to biased tracers of the matter density distribution, such as LRG. This effect was tested using \Nb~simulations \citep{noh09, mehta11}. As shown in PaperI and PaperII, reconstruction improves the measurement on the acoustic scale by about $40\%$. Therefore, in this paper we use the reconstructed BAO data from SDSS DR7 (hereafter BAO) unless otherwise specified. In practice, the BAO data measures the acoustic scale relative to some fiducial cosmology. This ratio of the acoustic scales is defined to be $\alpha$, which is given by
\begin{equation}
\alpha=\frac{D_V/r_s}{D^{\rm fid}_V/r_s^{\rm fid}}
\label{eq:alpha}
\end{equation}
where $D_V$ is the spherically averaged distance scale to the pivot redshift, $D_V(z) = (D^2_A(z)cz/H(z))^{1/3}$, $r_s$ is the sound horizon scale, and ``fid'' stands for the values in the fiducial cosmology WMAP7 \citep{komatsu11}: $H_0 = 70.2$ km/s/Mpc, $\Ob h^2 = 0.02255, \Om = 0.274, n_s = 0.968, \sigma_8 = 0.816$. The sound horizon for our fiducial cosmology is $r_s^{\rm fid} = 152.76$ Mpc. With reconstruction, we measure the distance to $z = 0.35$ to be $D_V (z = 0.35)(r_s^{\rm fid}/r_s) =  1356 \pm 25$ Mpc (See PaperI).

We use the Markov-Chain Monte Carlo code CosmoMC (\href{http://cosmologist.info/cosmomc/}{http://cosmologist.info/cosmomc/}) to compute the constraints on the cosmological parameters \citep{lewis02}. The original BAO routine in CosmoMC was replaced in order to use information from the probability distribution function of the BAO peak location $p(\alpha)$ from the $\chi^2$ fitting described in PaperII. Using $p(\alpha)$, we estimate the likelihood of the acoustic scale $\alpha$ that corresponds to the cosmological parameters at a given step in the Markov chain. After the Markov chains converge, CosmoMC outputs the posterior probability distribution for each of the cosmological parameters, given the observations. Table~\ref{tab:params} gives the mean values and the RMS errors of the cosmological parameters for various cosmological models. 

\begin{figure}
\centering
\includegraphics[width=84mm]{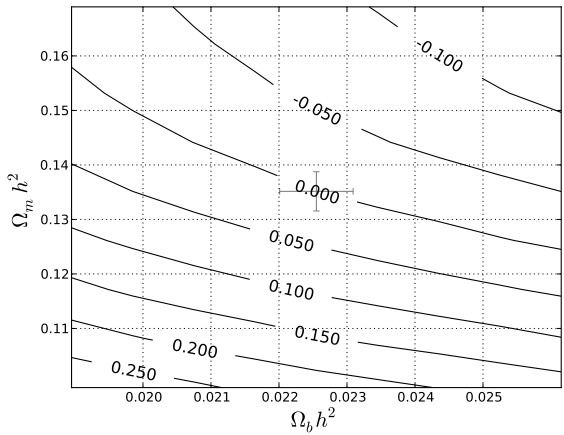}
\caption{Relative difference (in $\%$) between the sound horizon scale ($r_s/r^{\rm fid}_s)_{\rm CAMB}$ from CAMB and ($r_s/r^{\rm fid}_s)_{\rm EH98}$ from \citet{EH98}, for a given combination of ($\Om h^2, \Ob h^2$). Both definitions agree to within $0.2\%$ level even for cosmologies $5 \sigma$ away from the current WMAP7 constraints. Our fiducial cosmology with the WMAP7 1-sigma errors are shown as the grey cross.}
\label{fig:rs}
\end{figure}

A relevant issue that is often overlooked is that the sound horizon $r_s$ has several definitions in the literature. Its value depends on the definition of parameters such as $\zdrag$, the redshift at which the electrons are no longer dragged by the photons due to Compton scattering. In this paper, we use the definition of $\zdrag$ proposed in \cite{EH98} (hereafter EH98, Eq. 4 - 6) to compute the sound horizon, $r_s$. The difference between the definition of $r_s$ used in this paper and the implementation used in CAMB, is only a few percent for a wide range in $\Om$ and $\Ob$. In Figure~\ref{fig:rs} we show the relative difference of $r_s/r_s^{\rm fid}$ between the EH98 and CAMB definitions as a function of $\Om h^2$ and $\Omega_b h^2$, and find that the differences are negligible for our analysis. Thus the definition dependence cancels out when computing $\alpha$ except for a small residual $(\simeq 0.1\%)$ difference. 

In our CosmoMC chains, we use the WMAP7 data \citep{komatsu11} as our base dataset defined as ``CMB''. We then add our SDSS DR7 LRG reconstructed BAO data (PaperI and PaperII) to get the ``CMB+BAO'' dataset. We also include the other two latest BAO measurements from the 6-degree Field Galaxy Survey (6dFGS) \citep{beutler11} and the WiggleZ Dark Energy Survey \citep{blake11a}. The combination of WMAP7 and all BAO datasets is denoted by ``CMB+AllBAO''. While the SDSS and 6dFGS provide single redshift points, the WiggleZ survey measures BAO in three correlated redshift slices. We use all three redshift slices in our code and use their covariance matrix to account for the covariant points. \cite{conley11} provide a covariance matrix analysis of Type Ia supernova cosmology from the 3 year Supernovae Legacy Survey (SNLS3) accompanied by a CosmoMC module. We use their dataset and module in conjunction with the WMAP7 to create the ``CMB+SN'' dataset and add to our SDSS BAO data to create the ``CMB+BAO+SN'' dataset. Finally, we also use the direct $H_0$ measurement by \cite{riess11} and combine it with WMAP7, our SDSS BAO, and SNLS3 dataset into the ``CMB+BAO+H0+SN'' dataset. In the next section, we discuss the various cosmological parameters we measure using these datasets and how adding various datasets help measure and constrain various parameters in high dimensional cosmological models. 

\section{Results}\label{sec:results}

\subsection{Cosmology with BAO Data}\label{sec:ladder}
PaperI applies reconstruction to the SDSS DR7 LRG BAO dataset and PaperII shows the robustness of our BAO measurements. After using reconstruction, we measure $D_V (z = 0.35)(r_s^{\rm fid}/r_s) =  1356 \pm 25$ Mpc and  $D_V (z = 0.35)/r_s = 8.88 \pm 0.17$ giving us a $1.9\%$ measurement of the distance to $z = 0.35$. We can combine our BAO measurement with the measurements from 6dFGS \citep{beutler11} ($D_V = 456 \pm 27$ Mpc to $z = 0.106$) and WiggleZ Dark Energy Survey \citep{blake11} ($D_V = 2.23 \pm 0.11$ Gpc to $z = 0.6$) to make this BAO Hubble diagram. We have combined the three correlated WiggleZ redshift slices into one data point in order to show only uncorrelated points. 

\begin{figure}
\centering
\includegraphics[width=84mm]{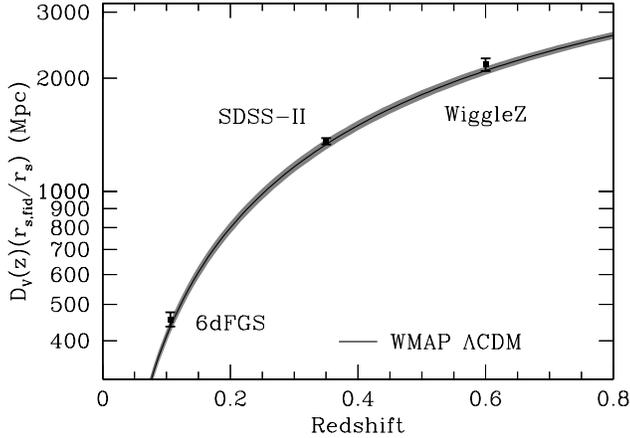}
\caption{6dFGS, reconstructed SDSS DR7, and WiggleZ BAO data points. The black line represents the $\LCDM$ prediction using WMAP7 data only \citep{komatsu11}. The shaded gray region is the effect of varying $\Om h^2$ within the $1 \sigma$ measurement errors of WMAP7. We see that the BAO data is consistent with the $\LCDM$ cosmological model.}
\label{fig:logDV}
\end{figure}

WMAP has measured the angular acoustic scale to about $0.1\%$ and has measured the baryon density with enough precision that its contribution to the sound horizon uncertainty is subdominant. Therefore, for any given value of $\Omega_m h^2$, we have a precise prediction for the sound horizon. Given an exact statement of the spatial curvature and $w(z)$, here flat $\LCDM$, only one value of $\Omega_m$ (and hence $H_0$) will satisfy the angular acoustic scale.  Hence, each value of $\Omega_m h^2$ makes a unique prediction for $D_V(z)/r_s$. We plot this prediction in Figure~\ref{fig:logDV} for the best-fit value of $\Omega_m h^2 = 0.1351$ as the solid black line, and the 1-sigma range of $\Omega_m h^2 = 0.1351 \pm 0.0051$ as the shaded region. This line is not a fit to the BAO data.

\begin{figure}
\centering
\includegraphics[width=84mm]{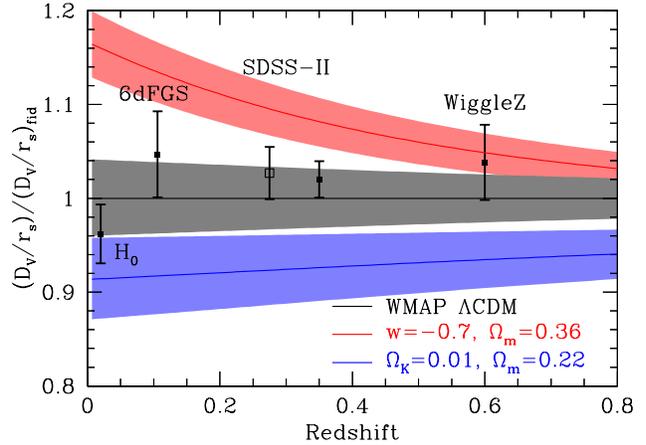}
\caption{Plot of $D_V/r_s$ normalized by the fiducial value. The open square is the \citet{percival10} BAO measurement. The black line is the WMAP7 $\LCDM$ model, red line shows the effect of varying $w$ and the blue line, the effect of varying $\Ok$. The shaded regions around these lines correspond to $1 \sigma$ uncertainty in $\Om h^2$ around the WMAP7 measurement. We see that the BAO data has the power to distinguish between various cosmological models. The $H_0$ point is the direct $H_0$ measurement from \citet{riess11}.}
\label{fig:DV}
\end{figure}

We see that the BAO data are a remarkable match to the WMAP7 $\LCDM$ prediction. To focus on the residuals, in Figure ~\ref{fig:DV}, we normalize the data to the WMAP7 best-fit model. Also plotted as the open square is the \cite{percival10} BAO data point at $z = 0.275$. Again, we see that the BAO data are consistent with $\LCDM$. As in Figure~\ref{fig:logDV}, for any assumption of $\Ok$ and $w(z)$, WMAP7 predicts a region on this plot with the width set by the uncertainty in $\Om h^2$. In this figure, we explore the effects of varying the equation of state parameter, $w$ and the curvature of the Universe $\Ok$ respectively. The red region corresponds to a flat Universe with $w = -0.7$, while the blue region corresponds to a Universe with a cosmological constant and $\Ok = 0.01$. $\Om$ is adjusted to keep the sound horizon constant. From this figure, we see that changing $w$ mostly changes the slope of the line on this plot while a non-zero $\Ok$ mostly changes the vertical offset. The relative distance measure from comparing the flux of SN constrain only the slope of the lines, while the BAO data can measure an absolute distance and hence the vertical offset.  This explains why SN data is more effective at constraining $w$, while the BAO data is more effective at constraining $\Ok$. The \cite{riess11} direct $H_0$ measurement is also plotted in this figure assuming the fiducial sound horizon value. While the sound horizon varies by about $1\%$ within the WMAP7 results, this effect is subdominant to the quoted errors on $H_0$. We explore the apparent tension between the BAO measurement and the direct measurement of $H_0$ in Section~\ref{sec:nrel}. 

Conventionally, the Hubble constant has been measured by building a distance ladder from local measurements out to measuring the cosmological Hubble flow. Conversely, the CMB and BAO data build an inverse distance ladder starting from a distance measurement at the recombination epoch. The CMB data provides an accurate measurement of the distance to the recombination redshift and our BAO data provides a measurement of distance to $z = 0.35$, thereby building an inverse distance ladder. The combination of these two datasets has the power to distinguish between different cosmological models. The supernovae data extrapolate the distance measurements to lower redshift and, therefore, precisely measure the expansion of the Universe at $z = 0$, which is the Hubble constant, $H_0$. In the following sections we use a combination of these datasets to explore a variety of cosmological models, and we use the CMB+BAO+SN dataset to obtain robust measurements of $H_0$ and $\Om$.

\begin{table*}
\caption{\label{tab:params} From top to bottom, the blocks correspond to the flat $\LCDM$ model (Section~\ref{sec:lcdm}), oCDM (Section~\ref{sec:ocdm}), $w$CDM (Section~\ref{sec:wcdm}), o$w$CDM (Section~\ref{sec:owcdm}), $w_0w_a$CDM (Section~\ref{sec:w0wacdm}), and o$w_0w_a$CDM (Section~\ref{sec:ow0wacdm}) respectively. The first column shows the data set used in each case, whereas the rest of the columns show the cosmological parameter values, with their uncertainties indicated in parenthesis. Empty values correspond to the cases in which the parameter is kept fixed to its fiducial value, i.e. $\Ok=0, w_0=-1, w_a=0$}

\begin{tabular}{lllllll}
\hline
Data Sets$^1$
 & $\Omega_{m} h^{2}$ & $\Omega_{m}$ & $H_{0}$ & $\Omega_{K}$ & $w_{0}$ & $w_{a}$ \\
& & & km/s/Mpc & & & \\
\hline

CMB & 0.1341(56) & 0.268(29) & 71.0(26) & ... & ... & ... \\
CMB+BAO & 0.1362(33) & 0.280(14) & 69.8(12) & ... & ... & ... \\
CMB+BAO+SN & 0.1349(33) & 0.274(14) & 70.2(12) & ... & ... & ... \\
\hline
CMB & 0.1344(55) & 0.423(175) & 60.0(123) & -0.039(44) & ... & ... \\
CMB+BAO & 0.1333(53) & 0.278(15) & 69.3(16) & -0.004(5) & ... & ... \\
CMB+AllBAO & 0.1326(50) & 0.277(13) & 69.2(14) & -0.004(5) & ... & ... \\
CMB+SN & 0.1324(51) & 0.243(37) & 74.6(58) & 0.003(9) & ... & ... \\
CMB+BAO+SN & 0.1323(50) & 0.274(13) & 69.6(16) & -0.004(5) & ... & ... \\
\hline
CMB & 0.1342(58) & 0.263(118) & 75.4(138) & ... & -1.12(41) & ... \\
CMB+BAO & 0.1349(57) & 0.285(25) & 69.0(39) & ... & -0.97(17) & ... \\
CMB+AllBAO & 0.1328(49) & 0.287(19) & 68.1(28) & ... & -0.92(13) & ... \\
CMB+SN & 0.1332(54) & 0.254(23) & 72.6(25) & ... & -1.04(7) & ... \\
CMB+BAO+SN & 0.1368(43) & 0.271(14) & 71.1(18) & ... & -1.05(8) & ... \\
\hline
CMB+AllBAO & 0.1321(51) & 0.281(30) & 68.9(39) & -0.001(10) & -0.97(24) & ... \\
CMB+SN & 0.1329(54) & 0.257(51) & 73.0(71) & 0.002(16) & -1.06(13) & ... \\
CMB+BAO+SN & 0.1336(52) & 0.271(14) & 70.3(19) & -0.005(5) & -1.08(8) & ... \\
CMB+BAO+SN+H0 & 0.1352(51) & 0.262(12) & 71.8(16) & -0.004(5) & -1.10(8) & ... \\
\hline
CMB+AllBAO & 0.1340(49) & 0.311(42) & 66.1(47) & ... & -0.62(47) & -0.88(122) \\
CMB+SN & 0.1345(53) & 0.242(24) & 74.7(31) & ... & -0.87(18) & -1.07(94) \\
CMB+BAO+SN & 0.1377(57) & 0.272(15) & 71.2(19) & ... & -1.02(16) & -0.26(82) \\
CMB+BAO+SN+H0 & 0.1385(55) &  0.266(14) & 72.2(16) & ... & -1.02(16) & -0.40(85) \\
\hline
CMB+AllBAO & 0.1327(50) & 0.303(46) & 66.7(51) & -0.003(11) & -0.66(47) & -1.11(122) \\
CMB+BAO+SN & 0.1346(53) & 0.276(15) & 69.9(19) & -0.010(7) & -0.90(16) & -1.30(99) \\
CMB+BAO+SN+H0 & 0.1363(53) & 0.267(13) & 71.4(16) & -0.008(6) & -0.94(16) & -1.23(102) \\
\hline
\end{tabular}
\newline
\begin{flushleft}
$^{1}$ {\footnotesize CMB = WMAP7, BAO = reconstructed SDSS DR7 LRG, SN = SNLS 3 year compilation, AllBAO = reconstructed SDSS DR7 LRG + 6dFGS + WiggleZ, H0 = \citet{riess11} measurement of $H_{0}$.}
\end{flushleft}
\label{tab:params}
\end{table*}

\subsection{$\LCDM$: The Vanilla Model}\label{sec:lcdm}
\begin{figure}
\centering
\includegraphics[width=84mm]{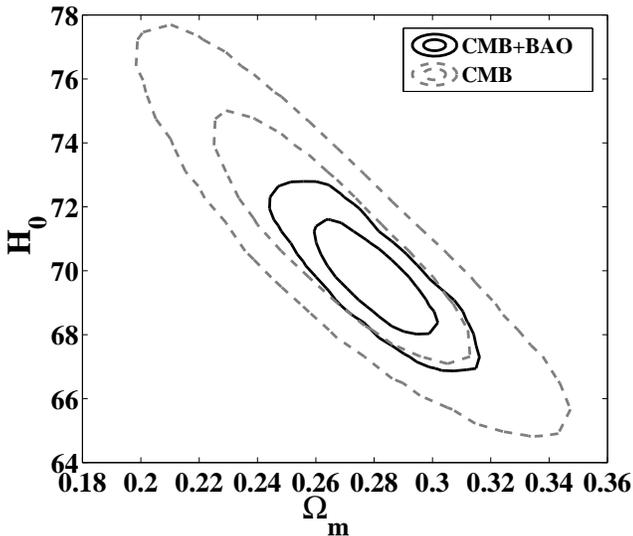}
\caption{$68\%$ and $95\%$ confidence level contours for $H_0$ vs $\Om$ using WMAP7 data (dashed gray lines) and then combining it with the reconstructed SDSS DR7 LRG BAO data (solid black lines).}
\label{fig:lcdm-omegamH0}
\end{figure}

The WMAP7 measurements of the CMB give us very good measurements of the various parameters in the ``vanilla cosmology'' model, also known as the $\LCDM$ model. In the CosmoMC code, we vary the standard CDM parameters of matter and baryon densities ($\Om, \Ob$), the primordial spectrum amplitude and slope ($n_s$), matter clustering amplitude ($\sigma_8$), and the optical depth to reionization ($\tau$).  Adding BAO measurement to the WMAP7 results improves the measurement of $\Om$ by about $40\%$ and $H_0$ by almost $30\%$. With reconstruction, we measure $\Om = 0.280 \pm 0.014$ and $H_0 = 69.8 \pm 1.2$ km/s/Mpc giving us a $1.7\%$ measurement of the Hubble constant. Figure~\ref{fig:lcdm-omegamH0} shows the $68\%$ and $95\%$ confidence level contours for $H_0$ vs $\Om$ and we can see the improvement in these parameters by adding the BAO data. Table~\ref{tab:params} shows the values for $\Omega_m h^2$, $\Omega_m$, and $H_0$ for various cosmological models and the corresponding datasets used. 

The acoustic standard ruler is calibrated by the WMAP measurement of $\Om h^2$. \cite{komatsu11} shows that allowing for a running spectral index, $dn_s/ d \ln k$ increases the errors on $\Om h^2$. Thus, we explore the effects of varying the running spectral index, $dn_s/d \ln k$ with the CMB and CMB+BAO datasets. We note that the nuisance parameters used in our BAO fitting techniques (PaperII) make our measurement of $D_V/r_s$ insensitive to the running spectral index. Table~\ref{tab:nrun} shows the effect of varying the running spectral index on cosmological parameters. We see that the running spectral index is consistent with 0: $d n_s/ d \ln k = -0.024 \pm 0.020$ using the CMB+BAO dataset. We find that including this parameter in the case of CMB data only, the $\Om h^2$ measurements are degraded by a factor of $1.4$ from $\Om h^2 = 0.1341 \pm 0.0056$ to $0.1393 \pm 0.0080$. This corresponds to an increased uncertainty in the measurements of $\Om$, $H_0$, and the spectral index $n_s$. Adding the BAO data improves the measurement of $\Om h^2$, $\Om$, $H_0$, and $n_s$ and are consistent with the values with no running spectral index. We also note that the value of $n_s$ is less than 1.0 in all cases which is expected by typical inflation models. 

\begin{table*}
\caption{\label{tab:nrun} Cosmological parameter values for a non-zero running of the spectral index in $\LCDM$. As in Table~\ref{tab:params}, the first column indicates the dataset used, whereas the rest of the columns indicate the measurements on the cosmological parameters. For comparison, rows 2 and 4 are taken from Table~\ref{tab:params}.}

\begin{tabular}{l l l l l l}
\hline
Data Sets \footnotemark[1] & $n_s$ & $dn_s/d\ln k$ & $\Omega_m h^2$ & $\Omega_m$ & $H_0$ \\
& & & & & km/s/Mpc \\
\hline
CMB & 0.929(40) & -0.028(27) & 0.1393(80) & 0.305(51) & 68.1(38) \\
CMB & 0.968(14) & ... & 0.1341(56) & 0.268(29) & 71.0(26) \\
CMB+BAO & 0.937(26) & -0.024(20) & 0.1372(36) & 0.289(17) & 69.0(14) \\ 
CMB+BAO & 0.964(12) & ... & 0.1362(33) & 0.280(14) & 69.8(12) \\
\hline
\end{tabular}
\newline
\begin{flushleft}
$^{1}$ {\footnotesize CMB = WMAP7, BAO = reconstructed SDSS DR7 LRG.}
\end{flushleft}
\label{tab:nrun}
\end{table*}

\subsection{oCDM: Varying Spatial Curvature}\label{sec:ocdm}
\begin{figure}
\centering
\includegraphics[width=84mm]{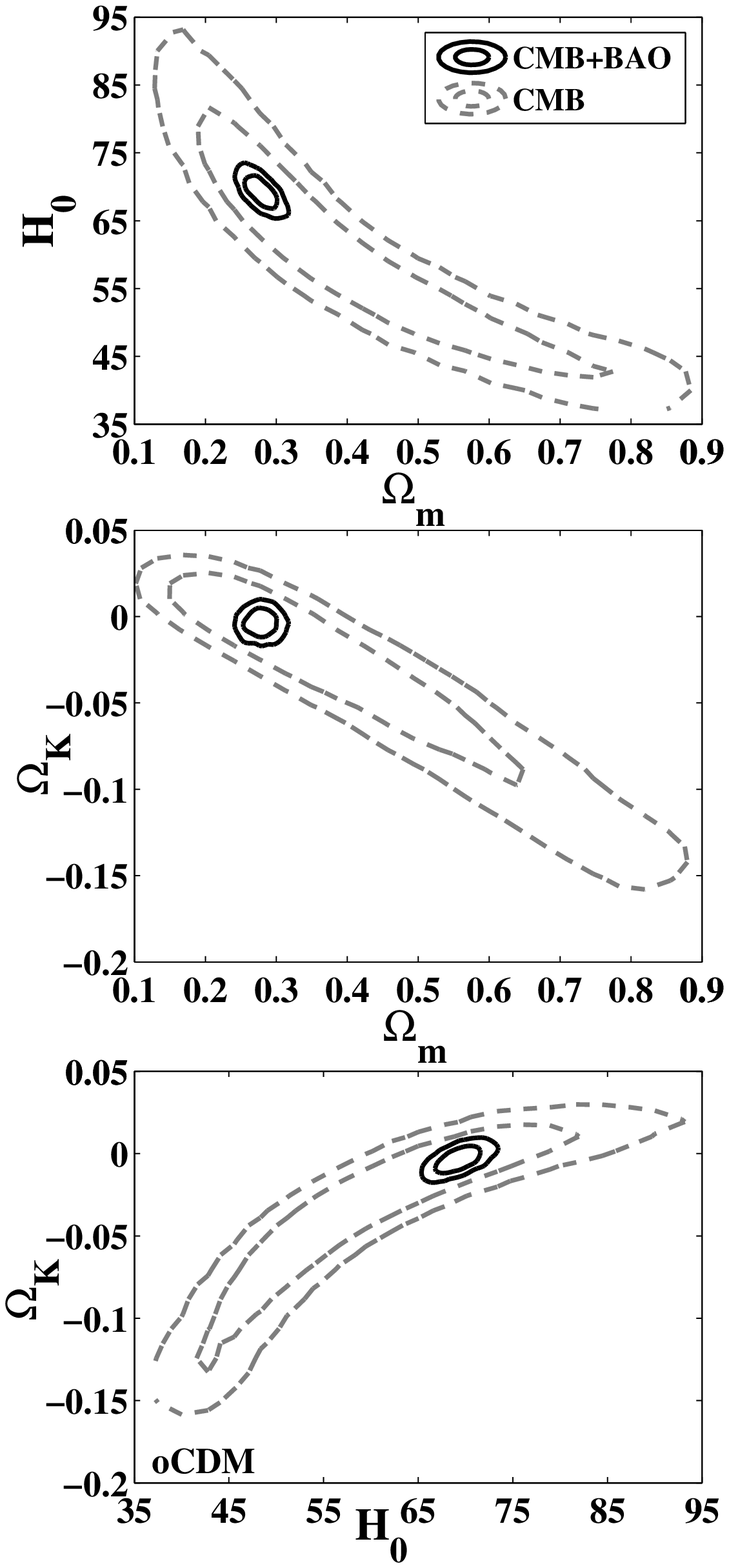}
\caption{$68\%$ and $95\%$ confidence level contours for $H_0$ vs $\Om$ (top), $\Ok$ vs $\Om$ (middle) and $\Ok$ vs $H_0$ (bottom) for the oCDM model. The gray dashed lines represent the ``CMB'' dataset, and the solid black lines represent the ``CMB+BAO'' dataset. We see the vast improvement in the parameter measurements by adding BAO data to the WMAP7 measurements.}
\label{fig:olcdm-2D}
\end{figure}

The BAO measurements calibrate the acoustic scale at low redshifts to the high redshift measurement from the CMB data. Therefore, the BAO accurately measures the curvature of the Universe (see Figure~\ref{fig:DV}). As shown in the previous section, in the $\LCDM$ model, the CMB breaks the $\Om - H_0$ degeneracy with a distance measurement to the recombination redshift. However, when we vary the curvature parameter $\Omega_K$ in the oCDM model, the CMB data has a degeneracy between $\Om$, $\Omega_K$, and $H_0$. The BAO measurement adds a second distance measurement in the inverse distance ladder, breaking this degeneracy and significantly improving these measurements. Using the BAO data, we find that the Universe is consistent with a flat geometry: $\Omega_K = -0.004 \pm 0.005$. Figure~\ref{fig:olcdm-2D} shows the improvements in the $H_0$ vs $\Om$, $\Omega_K$ vs $\Om$ and $\Omega_K$ vs $H_0$ contours. From these different panels and Table~\ref{tab:params}, we clearly see that the CMB degeneracies between these three parameters are greatly reduced by adding the BAO data.

\begin{figure}
\centering
\includegraphics[width=84mm]{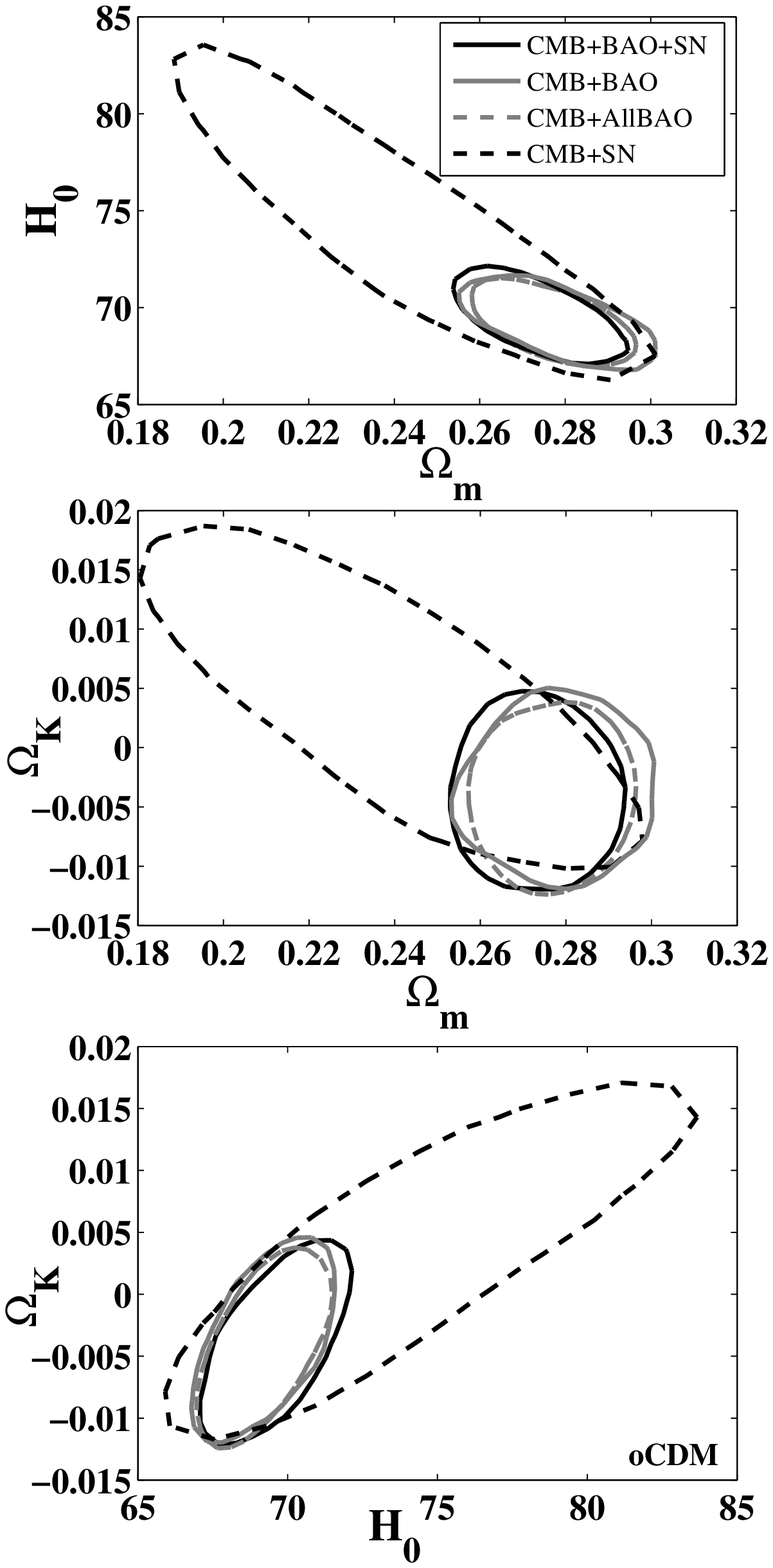}
\caption{$68\%$ confidence level contours for $H_0$ vs $\Om$ (top), $\Ok$ vs $\Om$ (middle) and $\Ok$ vs $H_0$ (bottom) for the oCDM cosmological model using the CMB+BAO+SN (solid black line), CMB+BAO (solid gray line), CMB+AllBAO (dashed gray line), and the CMB+SN (dashed black line) datasets.}
\label{fig:olcdm-compare}
\end{figure}

The ``AllBAO'' dataset gives us BAO measurements at various redshifts: $z = 0.106$ (6dF Galaxy Survey), $0.35$ (reconstructed SDSS DR7), $0.6$ (WiggleZ Dark Energy Survey). Figure~\ref{fig:olcdm-compare} shows the $68\%$ confidence level contours for various data sets. Table~\ref{tab:params} provides the values for the cosmological parameters. From the table and Fig.~\ref{fig:olcdm-compare}, we see that the SN and additional BAO data add little to constrain the parameters over the CMB+BAO dataset.

\subsection{$w$CDM: Varying the Constant Dark Energy Equation of State Parameter}\label{sec:wcdm}
\begin{figure}
\centering
\includegraphics[width=84mm]{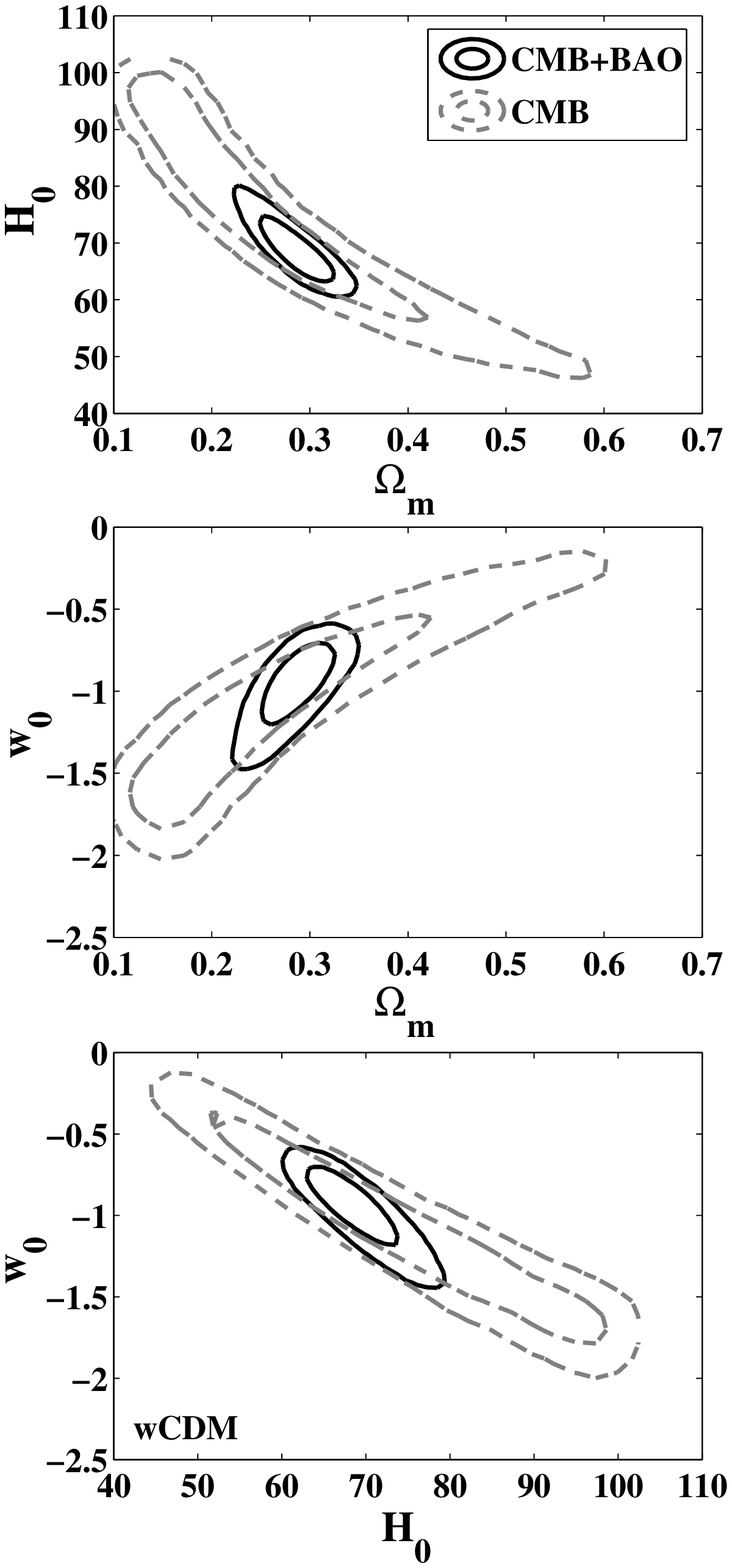}
\caption{$68\%$ and $95\%$ confidence level contours for $H_0$ vs $\Om$ (top), $w$ vs $\Om$ (middle) and $w$ vs $H_0$ (bottom) for the $w$CDM model. The grey dashed lines represent the ``CMB'' dataset, and the solid black lines represent the ``CMB+BAO'' dataset. We see the improvement in the parameter measurements by adding BAO data to the WMAP measurements.}
\label{fig:wcdm}
\end{figure}

In this section, we allow the dark energy equation of state parameter $w$ to vary and we measure its value. Using the CMB+BAO dataset, we measure the equation of state parameter $w = -0.97 \pm 0.17$, which consistent with a cosmological constant ($w = -1$). Figure~\ref{fig:wcdm} shows the $68\%$ and $95\%$ confidence level contour plots for $H_0$, $\Om$ and $w$. The combination of low redshift (SDSS DR7) and high redshift (WMAP7) measurement of the acoustic scale measures the expansion of the Universe and helps measure the equation of state parameter for dark energy. We see that the combined CMB and BAO data precisely measures $H_0$, $\Om$ and $w$ as listed in Table~\ref{tab:params}. Similar to the oCDM case (section~\ref{sec:ocdm}), we see that the CMB alone provides a robust measurement of $\Omega_m h^2$ but adding the BAO data breaks the degeneracy between the $H_0$, $\Omega_m$, and $w$. 

\begin{figure}
\centering
\includegraphics[width=84mm]{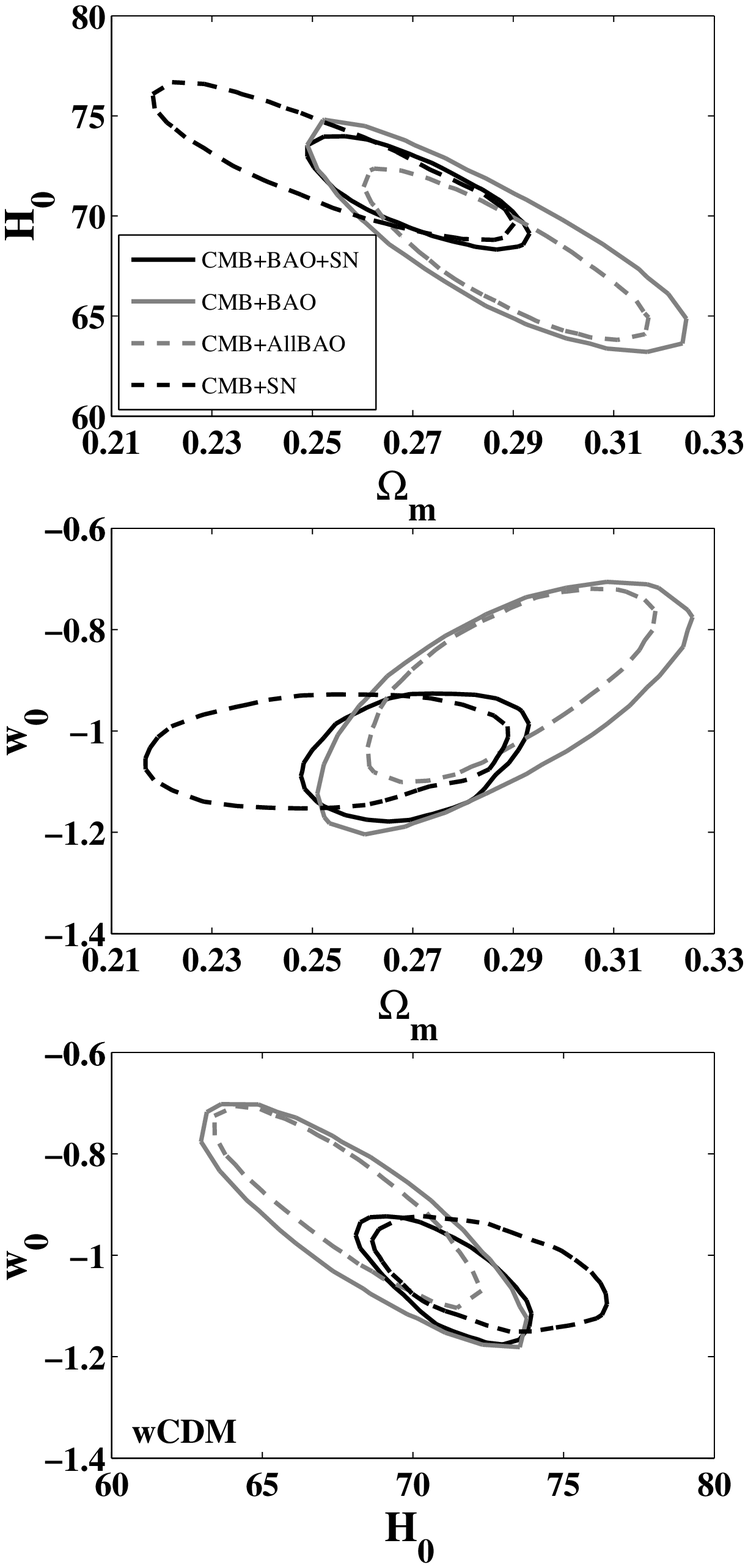}
\caption{$68\%$ confidence level contours for $H_0$ vs $\Om$ (top), $w$ vs $\Om$ (middle) and $w$ vs $H_0$ (bottom) for the $w$CDM model using the CMB+BAO+SN (solid black line), CMB+BAO (solid grey line), CMB+AllBAO (dashed grey line), and the CMB+SN (dashed black line) datasets.}
\label{fig:wcdm-compare}
\end{figure}

We compare these measurements of $H_0$, $\Omega_m$, and $w$ with values for different datasets. The CMB+AllBAO dataset slightly improves our measurements on $w$ and $H_0$ as shown in Table~\ref{tab:params}.  Figure~\ref{fig:wcdm-compare} shows the $68\%$ confidence level contours for $H_0$, $\Om$ and $w$. In the oCDM case, the BAO data precisely measured $\Ok$, $\Om$, and $H_0$ and we see no additional improvement by adding the SN data. In this case, however, we see that adding the SN data improves our measurements of $w$, $\Om$, and $H_0$. This is expected from Figure~\ref{fig:DV} as the BAO data constrains the vertical offset rather than the slope of the lines. We see that while all the measured values are consistent with each other within $1 \sigma$, the CMB+BAO dataset results tend to favor lower $H_0$ values and therefore higher values of $\Om$ and $w$ compared to the CMB+BAO+SN dataset. However, as Figure~\ref{fig:wcdm-compare} shows, the different data sets are consistent with the $\LCDM$ value of $w = -1$.

\subsection{o$w$CDM: Varying Curvature and the Constant Dark Energy Equation of State Parameter}\label{sec:owcdm}
\begin{figure}
\centering
\includegraphics[width=84mm]{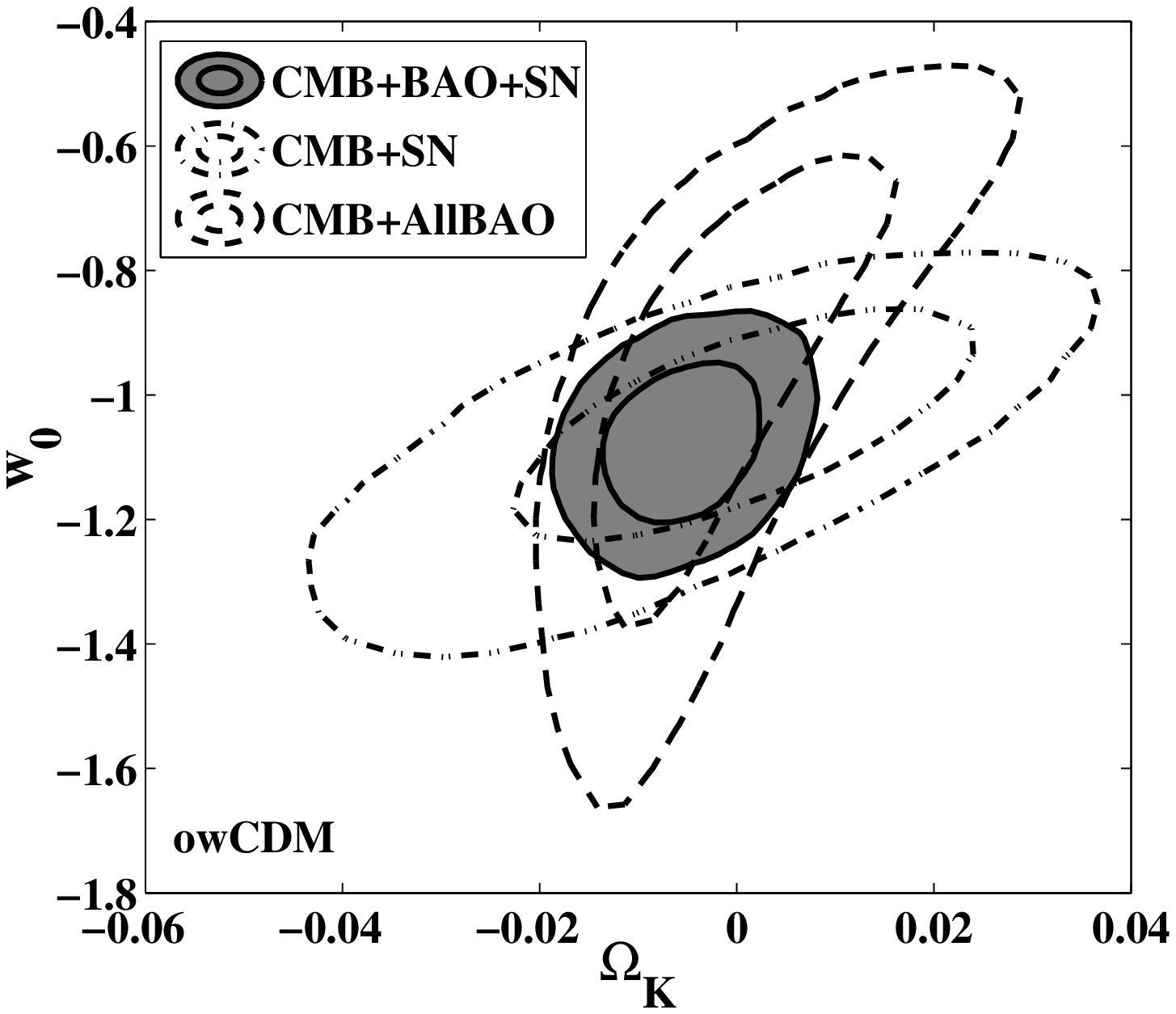}
\caption{$68\%$ and $95\%$ confidence level contours for $w_0$ vs $\Ok$ in the o$w$CDM cosmological model using the CMB+BAO+SN (solid lines with grey shading), CMB+SN (dot-dashed lines), and CMB+AllBAO (dashed lines) datasets. We see how the BAO and SN data, added to the CMB data, break the degeneracy between $w_0$ and $\Ok$.}
\label{fig:owcdm}
\end{figure}

Next, we move onto models where we vary two extra parameters in addition to the $\LCDM$ model parameters. In this case, we choose to vary both the dark energy equation of state parameter $w$, and the curvature parameter $\Omega_K$ as free parameters. Figure~\ref{fig:owcdm} shows the $68\%$ and $95\%$ confidence level contours for $w$ vs $\Omega_K$ for the CMB+AllBAO, CMB+SN, and CMB+BAO+SN datasets. We measure $w = -1.08 \pm 0.08$, and $H_0 = 70.3 \pm 1.9$ km/s/Mpc giving us an $4.4\%$ and $2.7\%$ measurement of $w$ and $H_0$. We precisely measure the curvature of the Universe to be consistent with being flat $\Omega_K = -0.005 \pm 0.005$. We note that the CMB+AllBAO results are consistent with the CMB+SN measurements. Adding the low redshift $H_0$ measurement by \cite{riess11} gives us consistent measurements with CMB+BAO+SN as shown in Table~\ref{tab:params}. 
 
\subsection{$w_0w_a$CDM: Varying the Time Dependent Dark Energy Equation of State Parameter}\label{sec:w0wacdm}
\begin{figure}
\centering
\includegraphics[width=84mm]{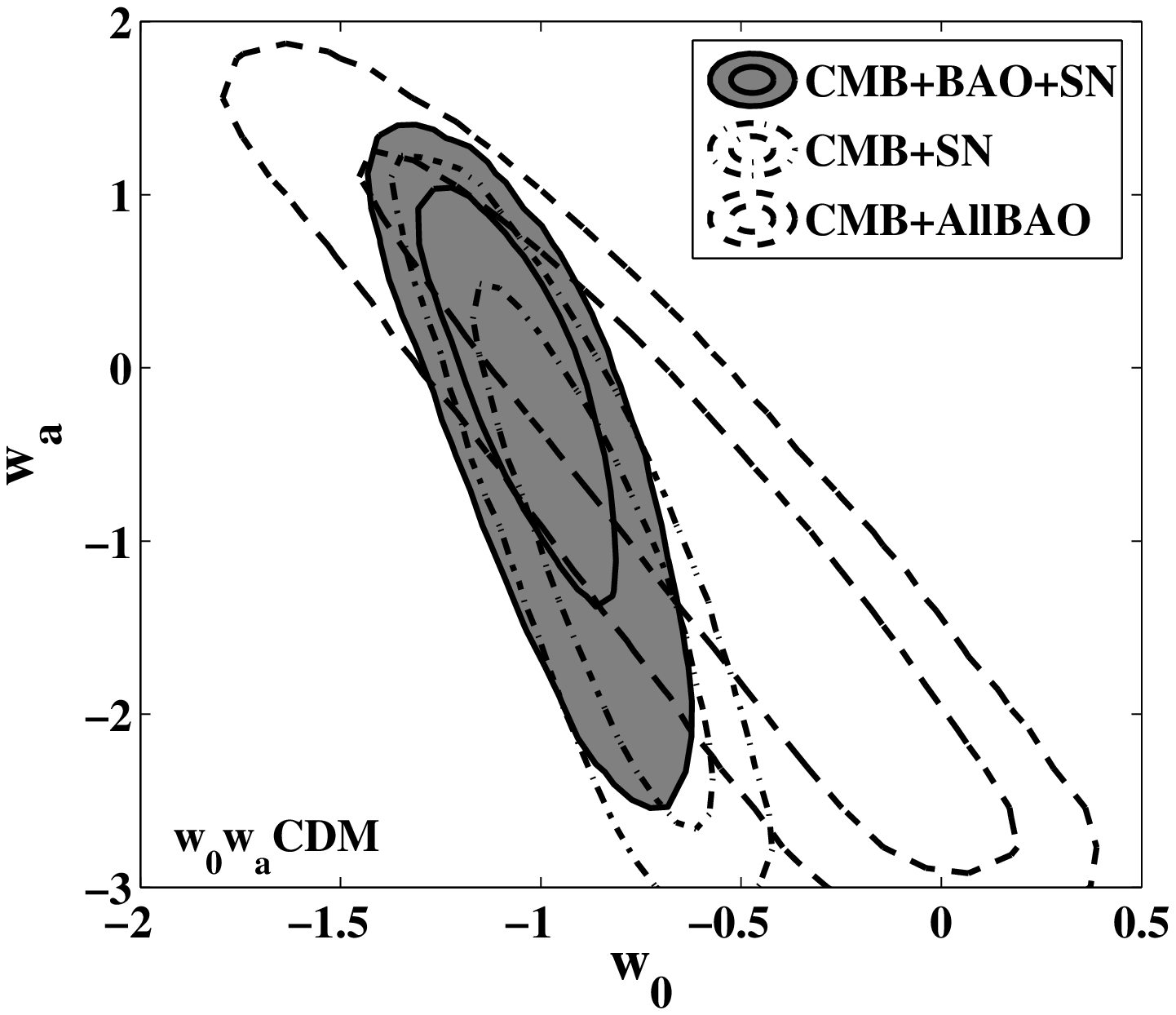}
\caption{$68\%$ and $95\%$ confidence level contours for $w_a$ vs $w_0$ in the $w_0w_a$CDM cosmological model using the CMB+BAO+SN (solid lines with grey shading), CMB+SN (dot-dashed lines), and CMB+AllBAO (dashed lines) datasets. We see that how the BAO and SN datasets provide different constraints in this parameter space.}
\label{fig:w0wacdm}
\end{figure}

As we probe higher redshifts, we can measure the evolution in $w$, the dark energy equation of state parameter. The most popular way to parametrize an evolving $w$, introduced by \cite{chevallier01} and \cite{linder03}, is:
\begin{eqnarray}
w(a) = w_0 + w_a(1-a) \Leftrightarrow \\ \nonumber
w(z) = w_0 + w_a\frac{z}{1+z},
\label{eq:wz}
\end{eqnarray}
where $a = 1/(1+z)$ is the scale factor. In this section, we assume a flat Universe and measure $w_0$ and $w_a$. For cosmological models that use this parameterization of dark energy equation of state ($w_0, w_a$),  we use the parameterized Post-Friedmann prescription for dark energy perturbations as implemented in a CAMB module \citep{lewis00} by Wenjuan Fang (\href{http://camb.info/ppf/}{http://camb.info/ppf/}). This modified code generalizes it to support a time-dependent equation of state $w(a)$. Figure~\ref{fig:w0wacdm} shows the $68\%$ and $95\%$ contours for $w_a$ vs $w_0$ using the CMB+BAO+SN, CMB+BAO, and CMB+AllBAO datasets. From Table~\ref{tab:params}, we see that we find very similar constraints on $H_0$ and $\Om$ as the previously presented cosmological models: $\Om = 0.272 \pm 0.15$ and $H_0 = 71.2 \pm 1.9$ km/s/Mpc giving us a $2.7\%$ measurement of $H_0$. We measure $w_0 = 1.02 \pm 0.16$ and $w_a = -0.26 \pm 0.82$, which are consistent with the $\LCDM$ model: $w_0 = -1$ and $w_a = 0$.
 
\subsection{o$w_0w_a$CDM: Varying Curvature and Time Dependent Dark Energy Equation of State Parameter}\label{sec:ow0wacdm}
\begin{figure}
\centering
\includegraphics[width=84mm]{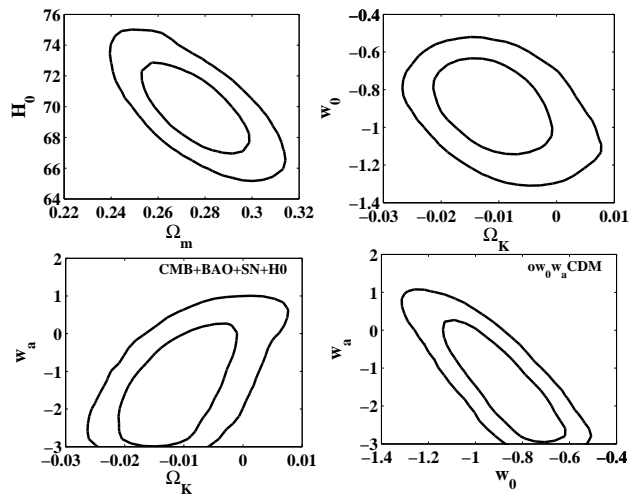}
\caption{$68\%$ and $95\%$ confidence level contours for $H_0$ vs $\Om$ (top left), $w_0$ vs $\Omega_K$ (top right), $w_a$ vs $\Omega_K$ (bottom left) and $w_a$ vs $w_0$ (bottom right) for the o$w_0w_a$CDM model using the CMB+BAO+SN dataset.}
\label{fig:ow0wacdm}
\end{figure}

In the most general cosmological model we analyze, we vary the curvature of the Universe, $\Omega_K$, and both the dark energy parameters $w_0$ and $w_a$ as free parameters. Figure~\ref{fig:ow0wacdm} shows the $68\%$ and $95\%$ contour levels for $H_0$ vs $\Om$, $w_0$ vs $\Omega_K$, $w_a$ vs $\Omega_K$, and $w_a$ vs $w_0$ using the CMB+BAO+SN dataset. It is noteworthy that even though we have curvature and both dark energy parameters as free parameters, the data is still consistent with a flat Universe with a cosmological constant. We see the precision in the measurements of $H_0$ and $\Om$ in the upper left panel of Figure~\ref{fig:ow0wacdm}. We obtain a $2.7\%$ measurement of $H_0 = 69.9 \pm 1.9$ km/s/Mpc and $w_0 = -0.90 \pm 0.16$. Our measurement of $w_a = -1.30 \pm 0.99$ is consistent with no evolution in $w(z)$. We measure $\Omega_K = -0.010 \pm 0.007$, which is still consistent with a flat Universe. We find that even with the high dimensionality of the cosmological model, we are able to measure and constrain various cosmological parameters.

The Dark Energy Task Force (DETF) compares various cosmology missions and defines their Figure of Merit (FoM) in the context of this cosmological model \citep{albrecht06}. The DETF FoM is defined as the inverse square-root of the determinant of the $w_0$-$w_a$ covariance matrix. The $68\%$ and $95\%$ contours for $w_a$ vs $w_0$ are shown in the bottom right panel in Figure~\ref{fig:ow0wacdm}. Using the CMB+BAO+SN dataset, we compute the DETF FoM to be 11.5. However, we note that this is an upper limit since the dataset allows $w_a$ outside our prior of $-3.0 \leq w_a \leq 2.0$. 

Table~\ref{tab:params} provides the values for CMB+BAO+SN, CMB+AllBAO, and CMB+BAO+SN+H0 datasets. We see that all the measured values are consistent with each other at the $1\sigma$ level. In order to prevent the Markov chains from exploring very extended and remote parameter spaces, we use a prior of $-3.0 \leq w_a \leq 2.0$. However, the chains run with CMB+BAO+SN and CMB+AllBAO datasets tend to allow values beyond $w_a < -3.0$ in their $95\%$ confidence level contours.

\subsection{Robust measurement of $H_0$ and $\Om$}\label{sec:H0Omegam}
\begin{figure}
\centering
\includegraphics[width=84mm]{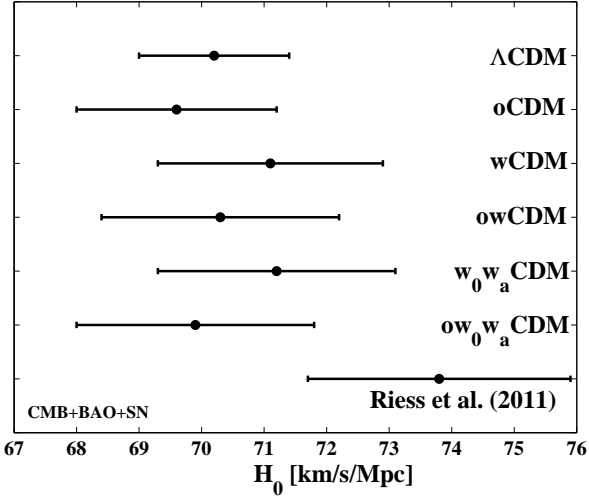}
\caption{Measurement of $H_0$ for various cosmological models with the CMB+BAO+SN dataset. Also included is the~\citet{riess11} $H_0$ measurement. We see that not only do we get a consistent and precise measurement of $H_0$ from CMB+BAO+SN dataset, this measurement is slightly lower than the $H_0$ measurement from the nearby Universe.}
\label{fig:H0Compare}
\end{figure}

\begin{figure}
\centering
\includegraphics[width=84mm]{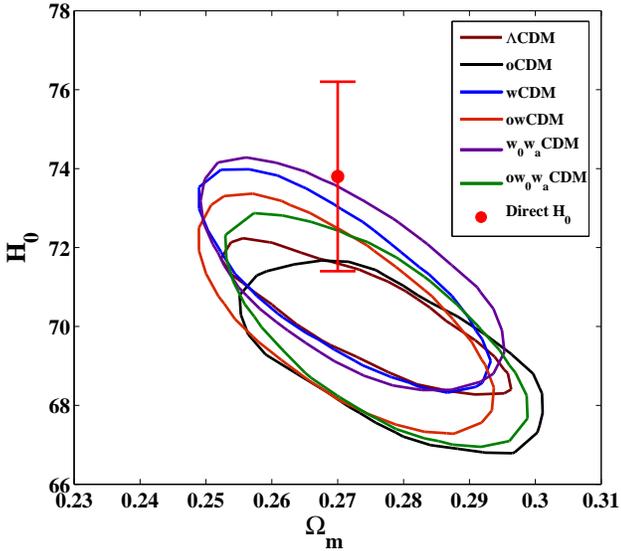}
\caption{$1 \sigma$ contours of $H_0$ in km/s/Mpc vs $\Om$ for various cosmological models using the CMB+BAO+SN dataset. Also plotted is the $H_0$ value measured by \citet{riess11}. We see that we get a robust measurement of $H_0$ and $\Om$ regardless of the cosmological model.}
\label{fig:H0Omegam}
\end{figure}

From previous sections, we have found that our measurements of $H_0$ and $\Om$ remain unchanged as we increase the dimensionality of our cosmological models. In this subsection, we explore this result and explain why our measurements of $H_0$ and $\Omega_m$ are robust regardless of the model for the late-time behavior of dark energy. In Figure~\ref{fig:H0Compare}, we show the measurements of $H_0$ from the CMB+BAO+SN datasets while varying the dark energy parameterization and the inclusion of spatial curvature. Figure~\ref{fig:H0Omegam} shows the same set of results as $1 \sigma$ contours in $H_0$ and $\Omega_m$. One can see that regardless of the cosmological model, we obtain highly consistent values and error bars for these quantities.

This robustness is due to the inverse distance ladder discussed in section~\ref{sec:ladder}. The CMB data provides a measurement of $\Omega_m h^2$ and the sound horizon $r_s$. The BAO data uses the measurement of $r_s$ to provide a distance measurement to $z=0.35$. The SN data then provide precise measures of the relative distance between $z=0.35$ and the local distance scale.  Hence, we have an empirical measure of the local distance and hence $H_0$, independent of spatial curvature or the model parameterization of the dark energy equation of state. Combining this measurement of $H_0$ with the CMB measurement of $\Omega_mh^2$ yields the value of $\Omega_m$.  We note that while this result is independent of the parameterization of late-time dark energy and the presence of spatial curvature, it would be sensitive to new cosmological physics at $z \geq 1000$ that alters the inference of $\Omega_m h^2$ and the sound horizon from the CMB.

It is of course important to compare our result for $H_0$ to direct measurements of the local distance scale. The Hubble constant has long been measured using distance ladders that build from local calibrations out to more distant galaxies situated in the Hubble flow \citep{freedman01, riess05, benedict07, riess09, freedman10}. A precise value of $H_0 = 73.8 \pm 2.4$ km/s/Mpc was recently obtained by the S$H_O$ES project \citep{riess11} using the NGC 4258 water maser \citep{argon07, humphreys08} and Cepheid variable stars measured in the near-infrared. We plot this measurement in Figure~\ref{fig:H0Compare} and \ref{fig:H0Omegam}. One sees that the direct measurement lies about $5\%$ higher than our inference from CMB+BAO+SN. However, this discrepancy only has a statistical significance of $1.5 \sigma$ and hence is not unusual. Nevertheless, we will return to this in the next subsection.

We note that the CMB+BAO+SN combination consistent favors $H_0$ values around 71, while CMB+BAO alone give slightly lower best-fit values of 69. The latter is not independent of the model for the expansion history; without SN, we are extrapolating the $z=0.35$ distance to $z\approx0$ using the cosmological model rather than an empirical measurement. Similarly, \citet{beutler11} measure $H_0 = 67.2 \pm 3.2$ km/s/Mpc using a BAO detection at $z=0.1$.  While this is all well within statistical uncertainties, apparently there is a small difference between the SN distance-redshift relation and that predicted from the combination of CMB and BAO data. 

\subsection{Energy Density of Relativistic Species}\label{sec:nrel}
The measurements of $H_0$ and $\Omega_m$ discussed in the previous section depend on knowledge of cosmological physics at $z\gtrsim 1000$. Further, we found a small tension between the CMB+BAO+SN measurement of $H_0$ and the direct measurement by \citet{riess11}. Hence, we are motivated to consider altering the standard cosmological model by adding additional relativistic particles with negligible interaction cross-section. These would be in addition to the usual cosmic background of the three neutrino species, and hence the new energy density is parameterized by altering the number of neutrino species from 3 to a new value $\Nrel$. We note that the particles need not actually be neutrinos, simply highly relativistic and negligibly interacting at late times. This possibility has a long history in cosmology, including constraints from Big Bang nucleosynthesis \citep{steigman77, hansen02, dolgov02, bowen02}. \citet{eisenstein04} pointed out that extra density in relativistic particles would cause CMB and BAO measurements to underestimate the value of $\Omega_mh^2$ and $H_0$. Numerous recent papers have constrained the density of relativistic particles with modern cosmology data \citep{seljak06, ichikawa07, mangano07, hamann10, reid10, riess11, giusarma11, komatsu11, calabrese11, archidiacono11}.

\begin{figure}
\centering
\includegraphics[width=84mm]{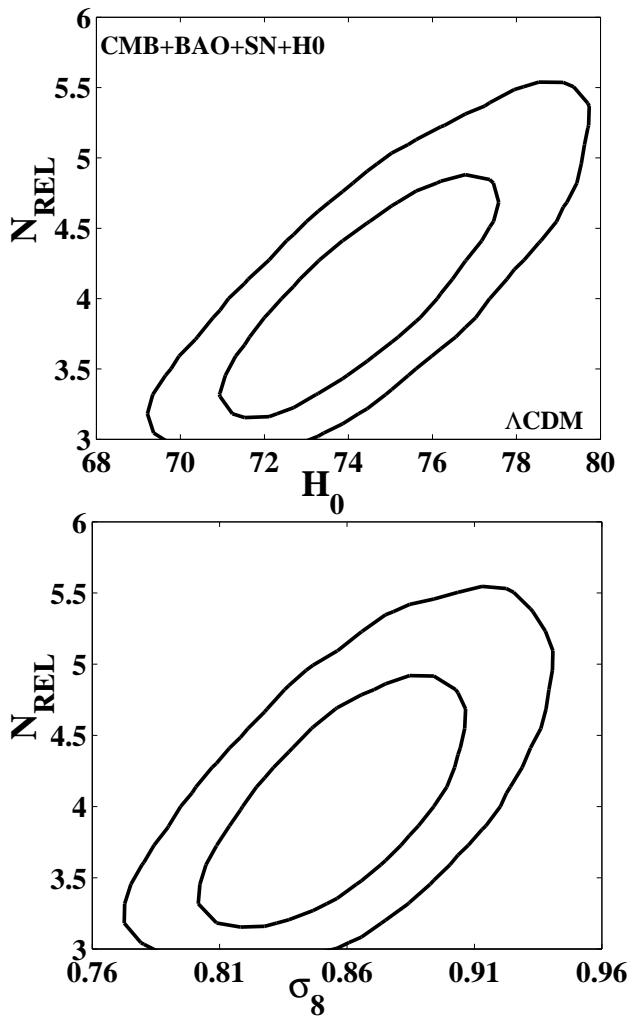}
\caption{$60\%$ and $95\%$ contours for $\Nrel$ vs $H_0$ (top panel) and $\Nrel$ vs $\sigma_8$ (bottom panel) using CMB+BAO+H0+SN dataset under $\LCDM + \Nrel$ cosmological model.}
\label{fig:nrel}
\end{figure}

\begin{table*}
\caption{\label{tab:nnu} Measuring the number of relativistic species, $\Nrel$, in the $\LCDM$, $w$CDM, and oCDM cosmological models.}

\begin{tabular}{l l l l l l l l}
\hline
Dataset \footnotemark[1] & $\Nrel$ \footnotemark[2] & $\Omega_m h^2$ & $\Omega_m$ & $H_0$  & $\sigma_8$ & $\Ok$ & $w$\\
& & & & km/s/Mpc & & &\\
\hline
CMB+BAO+H0+SN & 4.08(55) & 0.1524(104) & 0.275(13) & 74.5(21) & 0.86(3) & ... & ...\\
CMB+BAO+H0+SN & 4.00(58) & 0.1524(101) & 0.273(13) & 74.7(21) & 0.87(4) & ... & -1.03(8)\\
CMB+BAO+H0+SN & 4.22(60) & 0.1516(107) & 0.274(13) & 74.4(22) & 0.85(4) & -0.004(5) & ... \\
CMB+BAO+SN & ... & 0.1349(33) & 0.274(14) & 70.2(12) & 0.81(2) & ... & ... \\
CMB+BAO+SN & ... & 0.1368(43) & 0.271(14) & 71.1(18) & 0.84(4) & ... & -1.05(8) \\
CMB+BAO+SN & ... & 0.1323(50) & 0.274(13) & 69.6(16) & 0.80(3) & -0.004(5) & ... \\
\hline
\end{tabular}
\newline
\begin{flushleft}
$^{1}$ {\footnotesize CMB = WMAP7, BAO = reconstructed SDSS DR7 LRG, SN = SNLS 3 year compilation, H0 = \citet{riess11} measurement of $H_{0}$. $^{2}$ We use a prior of $\Nrel \geq 3.0$.}
\end{flushleft}
\label{tab:nrel}
\end{table*}

We therefore consider cosmological models that vary the relativistic density. In our MCMC chains, we use a prior of $N_{\rm REL} \geq 3$. Figure~\ref{fig:nrel} shows the $68 \%$ and $95\%$ confidence level contours for $\Nrel$ vs $H_0$ using the CMB+BAO+H0+SN dataset with a $\LCDM+\Nrel$ cosmology model. Table~\ref{tab:nrel} gives the values of $\Nrel$ and other cosmological parameters for three different models of the expansion history of the Universe. From Figure~\ref{fig:nrel} and Table~\ref{tab:nrel}, we see that the best-fit value for $\Nrel$ is around 4. Models with extra relativistic particle density increase the values of $\Om h^2$ and $H_0$, allowing a better fit to the \cite{riess11} measurement of $H_0 = 73.8 \pm 2.4$ km/s/Mpc. In terms of the inverse distance ladder, the added relativistic species affects $\Om h^2$, which moves the acoustic scale, and therefore changes the calibration of the distance ladder to larger values of $H_0$. Hence, it is not surprising to find that the other cosmological parameters such as $\Om$, $w$, and $\Ok$ remain unaffected by the addition of a new relativistic species. 

From Figure~\ref{fig:nrel}, we see that the shift away from $\Nrel = 3$ is not statistically significant. Table~\ref{tab:nrel} shows this shift to be about $2 \sigma$. This is larger than the $1.5 \sigma$ tension between the $H_0$ measurements; this is likely due to the $\Nrel \geq 3$ prior in our chains causing the mean value to be biased high and the variance to be biased low. However, a $2 \sigma$ shift when adding an extra parameter in our model is not compelling, but we note that recent cosmology results from the South Pole Telescope \citep{keisler11} and the Atacama Cosmology Telescope \citep{dunkley11} have found an excess of small-scale temperature anisotropy in the CMB, which could be explained by an extra density of relativistic particles beyond the usual neutrino background. 

This increase in the relativistic particle density also causes the model to predict a higher value of $\sigma_8$ as shown in the lower panel of Figure~\ref{fig:nrel} and in Table~\ref{tab:nrel}. The best-fit value shifts from $\sigma_8 = 0.81 \pm 0.02$ for $\Nrel = 3$ to $\sigma_8 = 0.86 \pm 0.03$ for $\Nrel \approx 4$. For comparison, \citet{allen11} (Table 2) gives a comparison of $\sigma_8$ measurements from galaxy cluster studies. X-ray \citep{henry09} and optical \citep{rozo10} studies of cluster abundances measure $\sigma_8 = 0.88 \pm 0.04$ and $0.80 \pm 0.07$, respectively. Thus, our $\sigma_8$ measurements for $\Nrel \approx 4$ are consistent with galaxy cluster measurements.

\section{Conclusions}\label{sec:conclusions}
In this series of papers, we have used the reconstructed SDSS DR7 LRG dataset to measure the BAO acoustic scale at the median redshift of $z = 0.35$. The reconstruction technique that provided this measurement has been discussed to great detail in Paper I, and the measurement itself has been extensively studied and tested in PaperII. In this paper, we use this BAO measurement of $D_V (z = 0.35)/r_s = 8.88 \pm 0.17$, which is a $1.9\%$ measurement of the distance to $z = 0.35$. To measure various cosmological parameters in a variety of cosmological models, we use our BAO data in combination with the CMB data from WMAP7 \citep{komatsu11} and the type Ia supernovae data from SNLS3 \citep{conley11} to extend the CMB+BAO inverse distance ladder to $z = 0$. With this CMB+BAO+SN dataset, we explore higher dimensional cosmological models and robustly measure the Hubble constant and the matter density of the Universe. We also use the BAO data from the 6dFGS and WiggleZ surveys in combination with our BAO data and the CMB data to measure cosmological parameters. In particular:

\begin{itemize}
\item We find that our BAO dataset is consistent with $\LCDM$ as shown in Figure~\ref{fig:logDV}. We improve on the WMAP7 measurements in $\LCDM$ and obtain a $1.7\%$ measurement of the Hubble constant: $H_0 = 69.8 \pm 1.2$ km/s/Mpc. 

\item As shown in Figure~\ref{fig:DV}, the distance measured from our BAO result is in good agreement with past work. It is therefore unsurprising that the cosmological parameters resulting from our chains are similar to those in recent works combining BAO with other data sets, e.g., \citet{komatsu09}, \citet{percival10}, \cite{reid10a}, \citet{blake10}, and \citet{beutler11}.

\item Paper I and PaperII show that reconstruction improves the BAO distance measurement by a factor of 1.8. We see this improvement as a reduction in the errors around $H_0$ and $\Omega_m$ by a factor of 1.5.

\item Under the $\LCDM$ model, we explore the effect of allowing a running spectral index, $d n_s /d \ln k$ and find that only using the CMB data degrades the measurements of $\Om h^2$ by a factor of 1.4. This translates into a larger uncertainty in measurements of $\Om, H_0$, and $n_s$. Adding BAO data decreases the uncertainty to 1.1. We also find that with both datasets that the value of $d n_s/d \ln k$ is still consistent with 0. 

\item The CMB+BAO dataset breaks the degeneracy between $H_0$, $\Om$, and $w$ or $\Ok$ in the $w$CDM and oCDM models respectively. We measure $w = 0.97 \pm 0.17$ and $\Ok = -0.003 \pm 0.005$, both consistent with $\LCDM$. We find that adding the other BAO data slightly improves the measurements on these parameters. 

\item For the higher dimensional cosmological models (o$w$CDM, $w_0w_a$CDM, and o$w_0w_a$CDM), we use the combined CMB+BAO+SN dataset to measure cosmological parameters. We find that even in these high dimensional models, the data is consistent with a flat Universe with a cosmological constant, i.e. consistent with $\LCDM$. The Dark Energy Task Force (DETF) Figure of Merit (FoM) is 11.5 using the CMB+BAO+SN dataset and using a prior on $w_a$.

\item Using the inverse distance ladder built from the CMB+BAO+SN dataset, we show that we obtain robust and precise measurements of both the Hubble constant and the matter density of the Universe despite varying the underlying model for the expansion history of the Universe. Even in our most general case, we measure $H_0 = 69.9 \pm 1.9$ km/s/Mpc and $\Om = 0.276 \pm 0.015$.

\item Our value of the Hubble constant is in mild tension (1.5 $\sigma$) with the direct measurement of $73.8 \pm 2.4$ km/s/Mpc by \cite{riess11}. We explore the possibility that this tension could be resolved by increasing the density of relativistic particles beyond the usual background of three species of neutrino.  We find that such a model can fit the $H_0$ value better if one adds density equivalent to 1 extra species of neutrinos. However, we stress that the conventional model is not rejected by our data. 

\end{itemize}

Looking towards the future, measurements of the distance-redshift relation with baryon acoustic oscillations will improve considerably. The SDSS-III Baryon Oscillation Spectroscopic Survey (BOSS) is underway and will extend the galaxy sample out to $z = 0.7$ \citep{eisenstein11}. We expect that the methods used in this SDSS DR7 analysis will be applicable to the BOSS sample. Yet larger surveys probing higher redshifts, such as Euclid and the Wide-Field Infrared Survey Telescope (WFIRST) missions, will use reconstruction to approach the cosmic variance statistical limit available to the acoustic peak method. We expect that baryon acoustic oscillations will play a major role in the precision mapping of the cosmic distance scale and expansion history of the Universe.

\section{Acknowledgments}
Funding for the Sloan Digital Sky Survey (SDSS) has been provided by the Alfred P. Sloan Foundation, the Participating Institutions, the National Science Foundation, the U.S. Department of Energy, the National Aeronautics and Space Administration, the Japanese Monbukagakusho, and the Max Planck Society, and the Higher Education Funding Council for England. The SDSS Web site is http://www.sdss.org/.

The SDSS is managed by the Astrophysical Research Consortium (ARC) for the Participating Institutions. The Participating Institutions are the American Museum of Natural History, Astrophysical Institute Potsdam, University of Basel, University of Cambridge, Case Western Reserve University, The University of Chicago, Drexel University, Fermilab, the Institute for Advanced Study, the Japan Participation Group, The Johns Hopkins University, the Joint Institute for Nuclear Astrophysics, the Kavli Institute for Particle Astrophysics and Cosmology, the Korean Scientist Group, the Chinese Academy of Sciences (LAMOST), Los Alamos National Laboratory, the Max-Planck-Institute for Astronomy (MPIA), the Max-Planck-Institute for Astrophysics (MPA), New Mexico State University, Ohio State University, University of Pittsburgh, University of Portsmouth, Princeton University, the United States Naval Observatory, and the University of Washington.

We thank the LasDamas collaboration for making their galaxy mock catalogs public. We thank Cameron McBride for assistance in using the LasDamas mocks, and comments on earlier versions of this work. We thank Martin White for useful conversations on reconstruction. We thank Bradford Benson for helping us find a bug in our code by pointing out a discrepancy between our constraints on the equation of state of dark energy using only CMB data and the reported values by the WMAP team. Finally, we thank the anonymous reviewer for helpful comments. KTM, DJE, and XX were supported by NSF grant AST-0707725 and NASA grant NNX07AH11G. NP and AJC are partially supported by NASA grant NNX11AF43G. The MCMC computations in this paper were run on the Odyssey cluster supported by the FAS Science Division Research Computing Group at Harvard University. This work was supported in part by the facilities and staff of the Yale University Faculty of Arts and Sciences High Performance Computing Center. 

\bibliographystyle{mn2elong}\bibliography{Kushal-MNRAS}

\clearpage

\end{document}